\documentclass[conference]{IEEEtran}
\usepackage{graphicx} 
\usepackage[utf8]{inputenc}
\usepackage{enumitem}
\usepackage[most]{tcolorbox}
\usepackage{nameref}
\usepackage{tikz}
\usetikzlibrary{automata, positioning, fit, arrows, shapes.geometric, arrows.meta, backgrounds}
\usepackage{footmisc}
\usepackage{svg}

\title{Employing LLMs for Incident Response Planning and Review}
\author{
    \IEEEauthorblockN{Sam Hays, Jules White}
    \IEEEauthorblockA{
        \textit{Department of Computer Science, Vanderbilt University, Nashville, TN, USA} \\
        \{george.s.hays, jules.white\}@vanderbilt.edu}
}
\date{January 2024}

\begin{document}

\maketitle

\section{Abstract}
\label{Abstract}
Incident Response Planning (IRP\footnote{IRP is also the abbreviation for ``incident response plan''.}) is essential for effective cybersecurity management, requiring detailed documentation (or \textit{playbooks}) to guide security personnel during incidents. Yet, creating comprehensive IRPs is often hindered by challenges such as complex systems, high turnover rates, and legacy technologies lacking documentation. This paper argues that, despite these obstacles, the development, review, and refinement of IRPs can be significantly enhanced through the utilization of Large Language Models (LLMs) like ChatGPT. By leveraging LLMs for tasks such as drafting initial plans, suggesting best practices, and identifying documentation gaps, organizations can overcome resource constraints and improve their readiness for cybersecurity incidents. We discuss the potential of LLMs to streamline IRP processes, while also considering the limitations and the need for human oversight in ensuring the accuracy and relevance of generated content. Our findings contribute to the cybersecurity field by demonstrating a novel approach to enhancing IRP with AI technologies, offering practical insights for organizations seeking to bolster their incident response capabilities.

\section{Introduction}
\label{Introduction}

In an era where digital threats evolve with increasing sophistication, the need for robust (cybersecurity) Incident Response Plans (IRPs) and appropriate Standard Operating Procedures (SOPs) has never been more critical. Incident Response Plans are the broad outline of what an organization may need to do to recover from various failures such as a malware infestation. 

An SOP is a guide to the actual steps needed to achieve certain outcomes for the given organization (e.g., secure an account, isolate equipment from the network). The NIST Contingency Planning Guide and the Computer Security Incident Handling Guide~\cite{Swanson_Bowen_Phillips_Gallup_Lynes_2010}~\cite{nist800_61r2} provide foundational frameworks for addressing cyber incidents, emphasizing the necessity for security teams to be well-prepared to mitigate and recover from attacks. This paper recognizes a gap in the current methodologies for developing IRPs and SOPs-specifically, the twin challenges of continuously updating these plans to address novel and esoteric attack vectors and the lack of composability of IRPs \& SOPs.

Responding effectively to cybersecurity incidents requires not only a deep understanding of potential threats but also creative and lateral thinking to anticipate and mitigate emerging vulnerabilities. Herein, we introduce a novel approach leveraging LLMs to enhance the design and planning of IRPs and SOPs. By incorporating LLMs, we aim to harness their capability for generating innovative solutions and continuous evaluation of existing plans against potential cyber threats.

Our contribution to the cybersecurity domain is in demonstrating how LLMs can facilitate the identification and creation of necessary response plans and procedures. We employ the ``SMART'' framework for defining meaningful constraints from which to work, and we illustrate the practical application of LLMs in cybersecurity planning, offering insights into their potential to significantly enhance organizational preparedness for security incidents.

\section{Incident Response Planning}
\label{IRP}
In this section we will discuss common steps for Incident Response Planning and the creation of SOPs for an organization. We will describe common areas of failure and demonstrate how LLMs can be leveraged to reduce likelihood or impact of those failures and improve the overall process. We also call attention to the notion that SOP creation is often out of scope during IR Planning, specifically SOPs are usually created by engineers or Individual Contributors (ICs) and IRPs are often created by management in tandem with legal departments or other higher-level personnel. We address this discontinuity and show how tight coupling of these tasks improves both.

\subsection{Traditional IRP}
\label{section:traditional_irp}
We base our definition of a traditional IRP largely on the NIST 800-61 Rev. 2~\cite{nist800_61r2} with addition perspectives from industry experience of the authors. 

The National Institute of Standards and Technology (NIST) outlines a structured approach for creating an incident response plan in ~\cite{nist800_61r2}. The process is divided into four main phases:

\begin{enumerate}
    \item \textbf{Preparation:} Establish and maintain an incident response capability. This includes training personnel, creating an incident response policy and plan, setting communication guidelines, and acquiring necessary tools and resources to improve organizational readiness for handling cybersecurity incidents~\cite{nist800_61r2}.
    \item \textbf{Detection and Analysis:} Focus on identifying and investigating suspected incidents through the analysis of precursors and indicators, gathering information from various sources, documenting incidents, and prioritizing them for response~\cite{nist800_61r2}.
    \item \textbf{Containment, Eradication, and Recovery:} Implement strategic decisions to contain the incident and prevent further damage, followed by the eradication of harmful elements from the system. The recovery process involves restoring systems or devices to their normal functions and monitoring to ensure the incident has been fully resolved~\cite{nist800_61r2}.
    \item \textbf{Post-Incident Activity:} Conduct a post-incident review to learn from the incident, prevent future occurrences, and improve the organization's incident response capability. This phase is essential for turning the incident into a learning opportunity to enhance future responses and strengthen the security posture~\cite{nist800_61r2}.
\end{enumerate}

The NIST guidance (and many other frameworks) describe the need to ensure a team is prepared to handle incidents and describes some example pieces of information and tools which might be necessary. For example: Contact Information for team members both within and external of the company, War Room area with shared knowledge of starting and operating it, Encryption software, etc. These items must be manually collected, made available, and maintained over time - even in the absence of their usage. Further, a long period of non-use will likely lead to \textit{software decay}\footnote{\label{software_decay}Software Decay or ``bit rot'' is a circumstance under which software becomes buggy or unusable over time through changes unrelated to the software itself, such as OS updates.}.

After the team's tools, procedures, and training is in a state with which everyone agrees that they has achieved the level of maturation adequate for handling ``live'' incidents, then phase 2 (Detection and Analysis) can be undertaken. In this phase, the guidance describes building an environment which is, firstly, resilient to attacks to the greatest degree possible. Secondly, ensuring detection rules are in place across the various technologies in use and, if at all possible, ensuring correlative analysis is possible for detecting both precursors (data suggestion an attack might be forthcoming) and actual indicators (indications that an attack has happened or is currently happening). 

When both detection is in place and the team is capable of handling it, phase 3 (Containment, Eradication, and Recovery) will start. This is the phase in which the IRPs are executed upon, SOPs might be employed, and ultimately the incident resolved. Within any framework, there is a significant gray area at this point because few incidents are exactly alike and many might include novel elements or combinations of attacks that are unique to an organization due to their specific set of hardware, software, cloud, and human characteristics.  Still, NIST does recommend that key classes of attack have specific IRPs ready to go. Examples include: Malware Infestation, Denial of Service Attacks, and other. 

Finally, after the incident has been resolved (phase 3 completed), then the post-incident activities can be scheduled and executed. These largely involve reviewing the incident, the actions taken, determining which things could have gone better, and making adjustments. 

There are some common challenges with this approach unrelated to the approach itself. The devil is in the details or, more correctly, in the maintenance of the plan. For example: if an IRP describes contacting Alice in the Legal department when an incident occurs and includes her contact information, then this is only accurate so long as Alice is a) Still employed, b) Available, c) Still the correct and responsible party.  If any of the conditions in that conjunction become false, then the IRP instructions become worse-than-useless because they are now misleading rather than just missing. 

Another example might be that the company has switched from one cloud provider to another for some common service, such as Single Sign-On (SSO). In this scenario, a large scale engineering project might have been undertaken over the course of months and the SSO environment was migrated. Now, the IRP might represent language associated with the previous service which would mislead incident responders who are new to the team or who had not been aware of the migration. It could also cause confusion when reading through documentation. Additionally, fidelity to the SOPs for specific types of actions might not have been written and consequently slow down the procedures (e.g., Instructions on resetting the MFA on account are now wrong; a severely problematic issue when time is of the essence during an incident). 

It is within this context, that human frailty and the protean nature of business objectives and prioritization, can cause IRPs and SOPs to become obsolete over the course of time, without themselves having changed (similar to software-decay\footref{software_decay}). 

\subsection{LLM-Augmented IRP}
We now juxtapose a new contribution to IRP: LLM-Augmented IRP. From Section~\ref{section:traditional_irp} we appreciate that there are many challenges to producing and maintaining IRPs and their associated SOPs which may not reflect negatively on any individual contributor. By leveraging LLMs in the ways we describe in this section, many of the common pitfalls can be overcome and a stronger IR Planning process, and SOP-building process can be achieved.

\subsubsection{IRPs and SOPs}
IRPs and SOPs have a relationship with each other which, while described in the NIST guidelines, is sometimes underrepresented in industry. NIST describes SOPs this way: \textit{``Procedures should be based on the incident response policy and plan. Standard operating procedures (SOPs) are a delineation of the specific technical processes, techniques, checklists, and forms used by the incident response team. [...] following standardized responses should minimize errors, particularly those that might be caused by stressful incident handling situations. SOPs should be tested to validate their accuracy and usefulness, then distributed to all team members''}. 

\begin{figure}
    \centering
    \begin{tikzpicture}[
        >={Latex[width=2mm,length=2mm]},
        every node/.style={align=center},
        irpnode/.style={rectangle, draw=black, fill=blue!20, minimum width=1cm, minimum height=0.5cm},
        sopnode/.style={ellipse, draw=black, fill=green!20, minimum width=.75cm, minimum height=1cm}
      ]
    
      \node[irpnode] (irp) {IRP\(_1\)};
      \node[sopnode, below left=of irp] (sop1) {SOP\(_1\)};
      \node[sopnode, below=of irp] (sop2) {SOP\(_2\)};
      \node[sopnode, below right=of irp] (sop3) {SOP\(_3\)};
    
      \draw[->] (irp) -- (sop1);
      \draw[->] (irp) -- (sop2);
      \draw[->] (irp) -- (sop3);
    
    \end{tikzpicture}
    \caption{Simplistic IRP-SOP Relationship}
    \label{fig:irpsop_rel_1}
\end{figure}
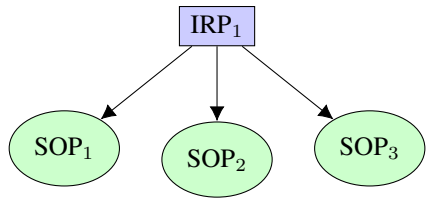

Notably, SOPs are typically the purview of an individual or team which is responsible for a specific technology, such as the process to reset an MFA token, or resetting a password. An SOP's relationship to an IRP can be visualized in Figure~\ref{fig:irpsop_rel_1}. An IRP is typically a larger-scale undertaking which includes management and is often built with input from across departments. The disconnect happens in the area where high-level IRP directions (``disconnect the computer from the network'') meet real-world application (``log into tool X with an account having security-controller rights, find the device in question, right-click [...]''). 

It is not uncommon for companies to either a) not have SOPs at all or b) for them to be out of date. In these scenarios, there is an implicit trust that the current staff will be able to handle whatever challenges come their way. This is often the case, but in situations of high-stress such as an incident response event, time is usually a factor and optimizing for the correct execution is critical. Indeed, hoping that an engineer or administrator performs all the right actions correctly and skips no steps, while simultaneously satisfactorily documenting their actions could be construed as, at best, unlikely and at worst, unreasonable.

Where LLMs can assist with this is in the planning and building stages of defining the SOPs that might be needed for a particular technology stack. By way of example, we provide a broad request to ChatGPT 4 and receive a response that is immediately helpful seen in Figure~\ref{fig:ad_sops}.

\begin{figure}
    \centering
    \begin{tcolorbox}[colback=gray!10,colframe=black,sharp corners]
        \small
        \textbf{Prompt:} ``You are helping me figure out which SOPs I might need to support various IRPs. The team I am on supports the Microsoft Active Directory environment and we want a list of SOPs that might be useful in any common security scenario. Please enumerate the titles and a brief description of recommended SOPs''\\
        \\
        \textbf{ChatGPT 4 Response:} ``Supporting a Microsoft Active Directory (AD) environment requires a comprehensive set of Standard Operating Procedures (SOPs) to efficiently manage security incidents. These SOPs should cover a wide range of scenarios, from user account management to the handling of more complex security breaches. Below is a list of recommended SOPs with brief descriptions for each, tailored to common security scenarios in an Active Directory environment:
        \begin{enumerate}
            \item \textbf{User Account Management}: Procedures for creating, modifying, disabling, and deleting user accounts, ensuring that access rights and privileges are in line with the principle of least privilege.
            \item \textbf{Password Management and Policies}: Guidelines for setting, updating, and managing passwords, including complexity requirements and rotation policies, to maintain account security.
            \item \textbf{Handling Stolen or Compromised Credentials}
            \item \textbf{Group Policy Object (GPO) Management}
            \item \textbf{Directory Services Restoration}
            \item \textbf{Security Group Management}
            \item \textbf{Monitoring and Responding to AD Alerts}
            \item \textbf{Patch Management for Domain Controllers}
            \item \textbf{Handling a Domain Controller Compromise}
            \item \textbf{Securing LDAP (Lightweight Directory Access Protocol) Communications}
            \item \textbf{Implementing Multi-Factor Authentication (MFA)}
            \item \textbf{Audit Logging and Review}
            \item \textbf{Disaster Recovery Planning}
            \item \textbf{Delegation of Administrative Privileges}
            \item \textbf{Secure Management of Service Accounts}''
        \end{enumerate}        
    \end{tcolorbox}
    \caption[Example LLM Prompt and Response]{Example LLM Prompt and Response\footnotemark}
    \label{fig:ad_sops}
\end{figure}
\footnotetext{Descriptions omitted for brevity after item 2.}

We see that based on very little context in the request, a set of 15 SOPs are recommended. We can further interrogate the LLM to provide more information as shown in Figure~\ref{fig:ad_prio_req_2}. This technique gives us actionable items against which to work and which the operational maturity can be measured (see section~\ref{smart:measurable}). 

\begin{figure}
    \centering
    \begin{tcolorbox}[colback=gray!10,colframe=black,sharp corners]
        \small
        \textbf{Prompt:} ``Based on your most current knowledge, please prioritize the list of SOPs and provide the top three most important ones to maximize effectiveness in incident response scenarios. Please provide a one sentence rationale for each choice.''\\
        \\
        \textbf{ChatGPT 4 Response:} ``Prioritizing SOPs for effectiveness in incident response within a Microsoft Active Directory (AD) environment, the top three, based on their immediate impact on security posture and incident mitigation capabilities, would be: 
        \begin{enumerate}
            \item \textbf{Handling Stolen or Compromised Credentials}: This SOP is critical because compromised credentials are a common attack vector and can lead to unauthorized access to sensitive systems and data, requiring swift action to mitigate risks.
            \item \textbf{Monitoring and Responding to AD Alerts} Effective monitoring and timely response to alerts are essential for early detection of potential security incidents, allowing for quick containment and minimizing damage.
            \item \textbf{Disaster Recovery Planning}: A well-defined disaster recovery plan ensures that the organization can quickly restore critical services in the event of a major incident, reducing downtime and operational impact.
        \end{enumerate}
        
        These priorities reflect the importance of proactive measures and rapid response in maintaining the security and integrity of Active Directory environments during incident response scenarios.''
    \end{tcolorbox}
    \caption{AD Prioritization Request}
    \label{fig:ad_prio_req_2}
\end{figure}

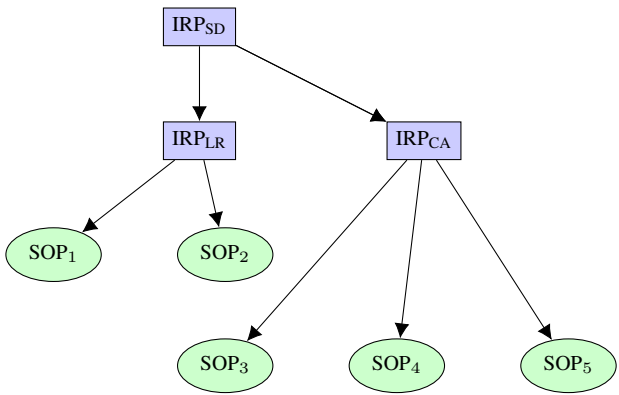
\begin{figure}[htbp]
    \centering
    \begin{tikzpicture}[
        >={Latex[width=2mm,length=2mm]},
        every node/.style={align=center},
        irpnode/.style={rectangle, draw=black, fill=blue!20, minimum width=0.2cm, minimum height=0.2cm, text depth=0.5ex, font=\footnotesize},
        sopnode/.style={ellipse, draw=black, fill=green!20, minimum width=0.2cm, minimum height=0.2cm, text depth=0.5ex, font=\footnotesize},
        node distance=1cm and 2cm 
      ]
      
      \node[irpnode] (irp1) {IRP\(_{\text{SD}}\)}; 
      
      \node[irpnode, below=of irp1] (intermediate1) {IRP\(_{\text{LR}}\)}; 
      \node[irpnode, right=of intermediate1] (intermediate2) {IRP\(_{\text{CA}}\)}; 
      
      \node[sopnode, below left=of intermediate1, xshift=1cm] (sop1) {SOP\(_1\)};
      \node[sopnode, right=of sop1, xshift=-1cm] (sop2) {SOP\(_2\)};
      \node[sopnode, below=of sop2, yshift=0.25cm] (sop3) {SOP\(_3\)};
      \node[sopnode, right=of sop3, xshift=-1cm] (sop4) {SOP\(_4\)};
      \node[sopnode, right=of sop4, xshift=-1cm] (sop5) {SOP\(_5\)};

      \draw[->] (irp1) -- (intermediate1);
      \draw[->] (irp1) -- (intermediate2);
      \draw[->] (irp1) -- (intermediate2);
      \draw[->] (intermediate1) -- (sop1);
      \draw[->] (intermediate1) -- (sop2);
      \draw[->] (intermediate2) -- (sop3);
      \draw[->] (intermediate2) -- (sop4);
      \draw[->] (intermediate2) -- (sop5);
      
    \end{tikzpicture}
    \caption{IRP-SOP Complex Relationship Diagram}
    \label{fig:irpsop_complex}
\end{figure}

When a Computer Security Incident Response Team (CSIRT) operationalizes their tasks in a general way, then various IRPs can be tightly coupled to the task (e.g., ``Secure the account'') but loosely coupled to the implementation. We see in Figure~\ref{fig:irpsop_complex} a series of interrelationships between IRPs and SOPs. In the diagram, each nodes represents a critical component in the Incident Response Planning (IRP) process, specifically tailored for managing incidents within an information technology framework. The nodes are defined as follows:

\begin{itemize}
    \item \textbf{IRP\(_{\text{SD}}\)}: Incident Response Plan for a Stolen Device. This plan outlines the steps to be taken in the event of a device theft, focusing on immediate response and mitigation strategies to secure potentially compromised information.
    \item \textbf{IRP\(_{\text{CA}}\)}: Incident Response Plan for a Compromised Account. This plan details the procedures for responding to incidents involving unauthorized access to user accounts.
    \item \textbf{IRP\(_{\text{LR}}\)}: Legal Response. A subset of IRP focused on the legal actions and compliance obligations triggered by a security incident.
    \\
    \item \textbf{SOP\(_1\)}: File Police Report. Standard Operating Procedure for reporting stolen devices to law enforcement agencies.
    \item \textbf{SOP\(_2\)}: Check Device Encryption. Procedure to verify the encryption status of a stolen or misplaced device to assess data vulnerability.
    \item \textbf{SOP\(_3\)}: Reset Password. Steps to promptly change passwords for accounts on the compromised or stolen device.
    \item \textbf{SOP\(_4\)}: Reset Multi-Factor Authentication (MFA). Guidelines for resetting MFA settings as a countermeasure against account compromise.
    \item \textbf{SOP\(_5\)}: Add Multi-Factor Authentication (MFA). Guidelines for adding an MFA token to an account as a countermeasure against compromise.
\end{itemize}

This methodology of building out SOPs that are appropriate for IRP allows a composable solution in which SOPs can be ``stacked'' in an IRP and thus ensures uniformity that the task is both well understood and handled consistently.

In Figure~\ref{fig:irpsop_complex}, for example, a Stolen Device IRP (\texttt{IRP\(_{\text{SD}}\)}) may reference all five SOPs and some or all of them may be executed, whereas a Compromised Account (\texttt{IRP\(_{\text{CA}}\)}) incident only references three SOPs (because a device was not stolen, a police remote may not be needed) and only some or all of those may be executed. 

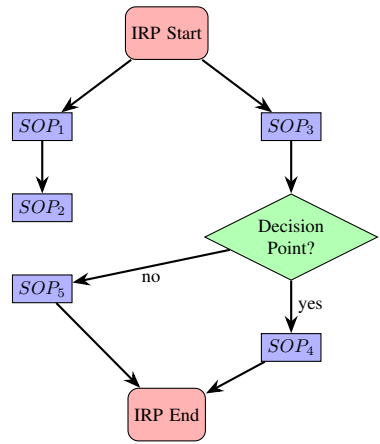
\begin{figure}[ht]
    \centering
    \begin{tikzpicture}[
        scale=0.7, transform shape, align=center,
        startstop/.style={rectangle, rounded corners, minimum width=0.5cm, minimum height=1cm,text centered, draw=black, fill=red!30},
        process/.style={rectangle, minimum width=0.5cm, minimum height=0.5cm, text centered, draw=black, fill=blue!30},
        decision/.style={diamond, aspect=2, minimum width=3cm, minimum height=1cm, text centered, draw=black, fill=green!30},
        arrow/.style={-Stealth, thick, black}
    ]    
        \node (start) [startstop] {IRP Start};
        \node (task1) [process, below right=of start] {$SOP_3$};
        \node (task2) [process, below left=of start] {$SOP_1$};
            \node (task4) [process, below=of task2] {$SOP_2$};
        \node (check) [decision, below=of task1] {Decision\\Point?};
            \node (task3) [process, below=of check] {$SOP_4$};
        \node (task5) [process, below=of task4] {$SOP_5$};
        \node (end) [startstop, below of=start, yshift=-6.25cm] {IRP End};

        \draw [arrow] (start) -- (task1);
        \draw [arrow] (start) -- (task2);
        \draw [arrow] (task1) -- (check);
        \draw [arrow] (check) -- node[anchor=west] {yes} (task3);
        \draw [arrow] (task2) -- (task4);
        \draw [arrow] (check) -- node[anchor=north] {no} (task5);
        \draw [arrow] (task5) -- (end);
        \draw [arrow] (task3) -- (end);
   
    \end{tikzpicture}
    \caption{Tabletop Exercise: Common Workflow}
    \label{fig:irpflowchart}
\end{figure}

An important distinction in the visualizations of Figures~\ref{fig:irpsop_complex} and \ref{fig:irpflowchart}: In Fig.~\ref{fig:irpsop_complex}, we see which playbooks are referenced by an IRP but not the conditions under which they may actually be invoked. On the other hand, Fig.~\ref{fig:irpflowchart} shows the actual flow of execution for SOP along with the branching logic that may exclude some of them. We see a mutual exclusion of \texttt{SOP\(_{\text{4}}\)} and \texttt{SOP\(_{\text{5}}\)} because of the decision point. 

These figures demonstrate the value in different types of visualizations, depending on the consumer. Figure~\ref{fig:irpsop_complex} provides a view of the directed graph of all referenced documents from a parent. This is useful in ensuring all documents are available and up to date. Figure~\ref{fig:irpflowchart} shows which circumstances will lead to the invocation of any one specific SOP (or possibly IRP), which is useful during the execution of the document.

\subsection{On Version Control}
\label{section:version_control}
An important concept for building an Incident Response Plan and supporting Standard Operating Procedures is that they do not exist in a vacuum. The environment will inevitably change around these documents and in so doing, they will become stale. To this end, and especially when augmented by an LLM, Semantic Versioning~\cite{semver} may be employed. Semantic Versioning is a versioning technique that makes clear the types of changes made to software, or in this case, documentation. 

In brief, a Semantic Version comes in three parts split by a period: MAJOR.MINOR.PATCH. The semantics are such that if the MAJOR part is incremented, then a breaking change has been introduced. If the MINOR or PATCH parts have been incremented, then a non-breaking change has been introduced.

This is valuable in documentation such as IRPs and SOPs because as an environment changes (e.g., new regulations are passed, new software is introduced or deprecated, new security controls are added, etc.), the documentation can reflect this. Taking an SOP as an example, we consider an SOP titled ``SOP: Reset MFA Token (1.0.0)''. At some point, a new employee notices that the user interface of the software does not reflect the screenshots of the documented procedure. They would then update the document and increment the MINOR or PATCH sections, depending on the size and scope of the change.  

In another case, if the company switched from Vendor A to Vendor B (e.g., Microsoft to DUO) for their MFA needs, then the entire document would need rewritten to reflect the current state of technology and the MAJOR version would need incrementing (indicating a breaking change). 

For the purposes of IRPs and SOPs, adding another section to the documentation can be helpful: MAJOR.MINOR.PATCH.REVIEWED. This last section may be used as a way to indicate that a document has been reviewed and is verified to be correct at that specific point in time and can be represented as an integer in the format of \texttt{MAJOR.MINOR.PATCH.[YEAR][MONTH][DAY]} (e.g., \texttt{1.2.14.20240107}), giving any reader an immediate sense of the document's freshness.
\section{Using LLMs to Build an IRP}
\label{section_example}
In this section we will describe patterns to use an LLM to build an IRP and describe the SOPs which will be needed to support it. To achieve these outcomes, we've adopt the ``SMART'' methodology~\cite{Dolan_1981} for setting and measuring the goal. The smart methodology ensures our goal is \textbf{S}pecific, \textbf{M}easurable, \textbf{A}chievable, \textbf{R}elevant, and \textbf{T}ime-bound. We will demonstrate that these characteristics are reasonable for building IRPs and SOPs with assistance from an LLM.

\subsection{Specific Goals}
When utilizing an LLM to assist with the construction of IRPs and SOPs, it is crucial to establish specific goals. This entails defining tasks in terms that are concrete and unambiguous. In instances where ambiguity exists, leveraging the LLM for clarification can be invaluable. The precision of the goal not only minimizes the likelihood of receiving noise or overly broad responses from the LLM but also ensures the system is appropriately bounded to focus on the specific problem at hand.

The security domain presents unique challenges in setting specific goals due to its broad cross-section of technology, risks, and compliance requirements. Nonetheless, by adequately setting the context for the LLM, it becomes feasible to start with a narrow focus, allowing the LLM to draw upon its extensive training data to provide relevant insights. Some of these insights may be directly actionable, while others might be disregarded based on their applicability to the specific context. 

\subsubsection{Effective Context Setting}
\begin{itemize}
    \item \textit{Contextualizing Prompts:} To achieve more targeted outputs from LLMs, it is essential to contextualize prompts effectively. This could involve specifying the technological environment (e.g., cloud-based infrastructure, on-premises data centers), detailing compliance standards (e.g., GDPR, HIPAA), or outlining known threat vectors relevant to the organization.
    \item \textit{Refinement Through Iteration:} Engaging with LLMs in an iterative process can help refine goals and resolve ambiguities. This iterative dialogue can include phased questioning or progressively detailed prompts based on initial LLM responses.
\end{itemize}

The work of trying various context breadth strategies to find a reasonable baseline is often a fruitful activity and can inform future starting points. A broad prompt might yield an overly general IRP, whereas a specific prompt, such as the one illustrated in Figure~\ref{fig:irp_generation}, ``[...] for a stolen device [...] outline an IRP and identify SOPs [...],'' leads to a more tailored and actionable response.

The quality and effectiveness of the resulting IRPs and SOPs are directly influenced by the specificity of the initial goals. Specific goals lead to IRPs and SOPs that are more aligned with the organization’s specific risk profile and operational needs, thereby enhancing their applicability and effectiveness in mitigating security risks.

Incorporating a feedback loop in the goal-setting process ensures that the development of IRPs and SOPs remains dynamic and responsive. Initial outputs from the LLM should be evaluated by key personnel within the organization, and the insights gained used to further refine and specify goals, thereby optimizing the relevance and applicability of the LLM-generated content.

\begin{figure}
    \centering
    \begin{tcolorbox}[colback=gray!10,colframe=black,sharp corners]
    \small
    \textbf{Prompt}: ``Let's build an Incident Response Plan for a stolen device (laptop, phone, etc.). Let's outline the IRP and then identify any SOPs that might be needed to support it. Note we use Bitlocker, Active Directory, Windows Laptops, and iPhone or Android mobile devices.''
    \\
    \textbf{Response}: ``\textbf{Incident Response Plan Outline for Stolen Device}
    \\
        \textbf{1. Incident Detection and Reporting}
        \textbf{Description:} Procedures to identify and report a stolen device. \\
        \textbf{Actions:}
        \begin{itemize}
            \item Employees must report the theft immediately to the IT department or designated security personnel.
            \item Use a form or email template to capture essential details (device type, last known location, data sensitivity, etc.).
        \end{itemize}
        
        \textbf{2. Initial Assessment}
        \textbf{Description:} Evaluate the incident's impact and classify its severity. \\
        \textbf{Actions:}
        \begin{itemize}
            \item Determine the type of data stored or accessed by the device.
            \item Assess the potential risk to the organization (data breach, financial loss, etc.).
        \end{itemize}
        
        \textbf{3. Containment}
        \textbf{Description:} Steps to prevent further data loss or unauthorized access. \\
        \textbf{Actions:}
        \begin{itemize}
            \item Remotely lock the device if possible.
            \item Use Active Directory to disable the user account associated with the device.
            \item Initiate remote wipe procedures for sensitive data if recovery seems unlikely.
        \end{itemize}
        
        \textbf{4. Eradication}
        \textbf{Description:} Ensure that no residual risk remains. \\
        \textbf{Actions:}
        \begin{itemize}
            \item Ensure that all sensitive data on the device is encrypted (via Bitlocker for Windows and respective encryption methods for mobile devices).
            \item Change all passwords for accounts accessed from the device.
        \end{itemize}
        
        \textbf{5. Recovery}
        \textbf{Description:} Steps to return to normal operations and prevent future incidents. \\
        \textbf{Actions:}
        \begin{itemize}
            \item Issue a replacement device to the user.
            \item Restore data from backups if necessary.
            \item Monitor for any unusual activity related to the stolen device.
        \end{itemize}
        
        \textbf{6. Post-Incident Analysis}
        \textbf{Description:} Review the incident to improve future response and prevent recurrence. \\
        \textbf{Actions:}
        \begin{itemize}
            \item Document the incident's details, response effectiveness, and any lessons learned.
            \item Update policies, procedures, and technical controls as needed.''
        \end{itemize}

    \end{tcolorbox}
    \caption{IRP Generation}
    \label{fig:irp_generation}
\end{figure}

\subsection{Measurable}
\label{smart:measurable}
\begin{figure}[h!]
    \centering
    \includegraphics[width=0.7\linewidth]{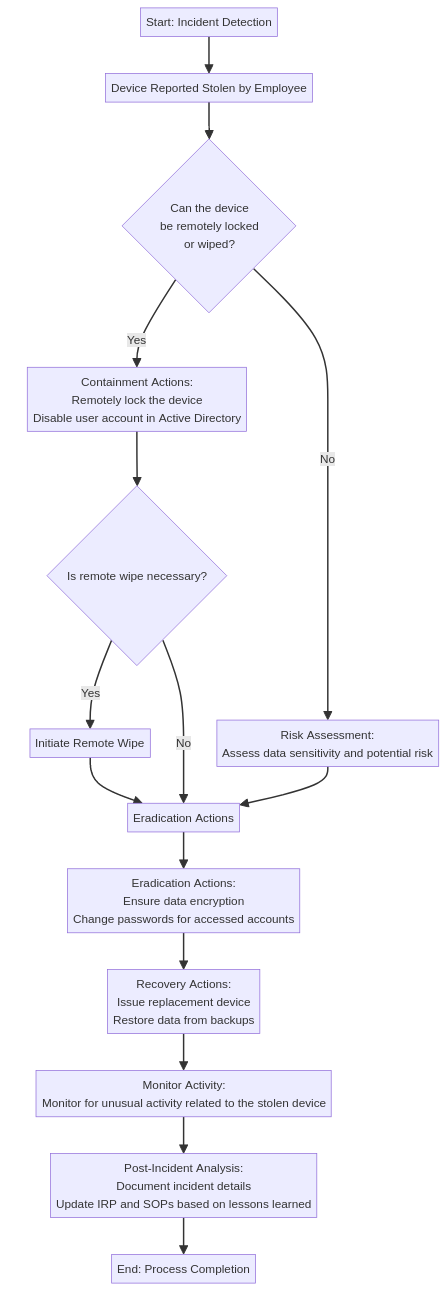}
    \begin{tcolorbox}[colback=gray!10,colframe=black,sharp corners]
        \small
        \textbf{Prompt:} Please render this flowchart using the mermaid language
    \end{tcolorbox}
    \caption{Mermaid diagram generated by ChatGPT}
    \label{fig:irp_mermaid}
\end{figure}

Measurement of the output of an LLM-assisted IRP or SOP can feel imprecise if the output is not of a mathematical nature. However, it can be strengthened by requesting flowcharts or other types of diagrams, such as state-machines, where appropriate. We see in Figure~\ref{fig:irp_mermaid} that ChatGPT was able to produce a flowchart representing the output of the request.

\begin{figure}[ht]
    \centering
    \begin{tikzpicture}[->, >=stealth', auto, semithick, node distance=1.2cm]
    \tikzstyle{every state}=[fill=white,draw=black,thick,text=black,scale=1,minimum size=10pt]
    \node[state, initial] (S1) {$S_{1}$};
    \node[state] (S2) [below of=S1] {$S_{2}$};
    \node[state] (S3) [below of=S2] {$S_{3}$};
    \node[state] (S4) [below left of=S3, xshift=-1.5cm] {$S_{4}$};
    \node[state] (S5) [below right of=S3, xshift=1.5cm] {$S_{5}$};
    \node[state] (S6) [below of=S4] {$S_{6}$};
    \node[state] (S7) [left of=S6, xshift=-2cm] {$S_{7}$};
    \node[state] (S8) [below of=S6] {$S_{8}$};
    \node[state] (S9) [below of=S8] {$S_{9}$};
    \node[state] (S10) [below of=S9] {$S_{10}$};
    \node[state] (S11) [below of=S10] {$S_{11}$};
    \node[state, accepting] (S12) [below of=S11] {$S_{12}$};
    
    \draw   (S1) -- (S2)
            (S2) -- (S3)
            (S3) -- node[above left] {Yes} (S4)
            (S3) -- node[above right] {No} (S5)
            (S4) -- (S6)
            (S6) -- node[above] {Yes} (S7)
            (S6) -- node[left] {No} (S8)
            (S5) -- (S8)
            (S7) -- (S8)
            (S8) -- (S9)
            (S9) -- (S10)
            (S10) -- (S11)
            (S11) -- (S12);
    \end{tikzpicture}
\textbf{State Descriptions:}
\begin{itemize}
    \item $S_{1}$: Incident Detection
    \item $S_{2}$: Device Reported Stolen by Employee
    \item $S_{3}$: Can the device be remotely locked or wiped?
    \item $S_{4}$: Containment Actions: Remotely lock the device, Disable user account in Active Directory
    \item $S_{5}$: Risk Assessment: Assess data sensitivity and potential risk
    \item $S_{6}$: Is remote wipe necessary?
    \item $S_{7}$: Initiate Remote Wipe
    \item $S_{8}$: Eradication Actions: Ensure data encryption, Change passwords for accessed accounts
    \item $S_{9}$: Recovery Actions: Issue replacement device, Restore data from backups
    \item $S_{10}$: Monitor Activity: Monitor for unusual activity related to the stolen device
    \item $S_{11}$: Post-Incident Analysis: Document incident details, Update IRP and SOPs based on lessons learned
    \item $S_{12}$: Process Completion
\end{itemize}
    \caption{FSM of Incident Response Process}
    \label{fig:incident_response_fsm}
\end{figure}

Flowcharts can often be turned into Finite State Machines for potentially easier analysis and completeness tests. The flowchart of Figure~\ref{fig:irp_mermaid} has been converted into a state machine in Figure~\ref{fig:incident_response_fsm} by ChatGPT. This can make using a formal solver, such as Z3 or others, easily available for more rigorous analysis. It is, however, unlikely that this specific level of rigor will be needed for SOPs or IRPs because they are \textit{usually} small enough for a human to reason about. Nevertheless, it is useful to have these techniques at our disposal when designing for Incident Response.

\subsection{Achievable}
In generating an IRP or SOP, especially in cases where the boundaries of the problem statement or the tools available are unclear, having a discussion with the internal resources of the company and the LLM can be useful. In cases where all variables are known, achieveability may not be possible due to competing resources. For this section we assume the resource contention problems are not considerations and instead focus on the achieveability of the IRP or SOP themselves.

\begin{figure}
    \centering
    \begin{tcolorbox}[colback=gray!10,colframe=black,sharp corners]
    \small
    \textbf{Prompt}: ``I would like to build an Incident Response Plan for an unknown person entering the building. We do not keep the office locked and have no security guard or security cameras or services. Is this possible? If not, what options have I?
    \\
    \textbf{Response}: 
\begin{enumerate}
    \item \textbf{Preparation}
    \begin{itemize}
        \item Assess Risks: Evaluate the likelihood and potential impact of unauthorized entry.
        \item Awareness and Training: Educate staff about potential risks and response strategies.
    \end{itemize}
    
    \item \textbf{Detection and Analysis}
    \begin{itemize}
        \item Visual Detection: Rely on employees to notice and report the presence of unknown individuals.
        \item Physical Sign-In: Implement a sign-in sheet at entry points for all visitors.
        \item Secure Sensitive Areas: Ensure that sensitive areas within the building are secured.
    \end{itemize}
    
    \item \textbf{Containment}
    \begin{itemize}
        \item Immediate Actions: Designate a safe area for employees and establish engagement protocols.
        \item Communication Plan: Develop a clear plan for internal and external communication.
    \end{itemize}
    
    \item \textbf{Eradication and Recovery}
    \begin{itemize}
        \item Remove the Threat: Involve appropriate measures to safely remove the threat.
        \item Learn and Adapt: Review the incident to improve future security measures and responses.
    \end{itemize}
    
    \item \textbf{Post-Incident Activity}
    \begin{itemize}
        \item Documentation: Record details of the incident for future reference and legal purposes.
        \item Review and Improve: Regularly update the IRP to address new threats and incorporate lessons learned.
    \end{itemize}
\end{enumerate}
    \end{tcolorbox}
    \caption{IRP Achievability Example (Truncated for brevity)}
    \label{fig:unknown_person}
\end{figure}

In cases where an IRP or SOP may seem impossible due to circumstances beyond one's control, the LLM may yet provide surprising and helpful guidance and, importantly, an outline of an achievable plan that either meets the need directly, or exposes areas where improvements must be made to achieve the goal to a higher degree of satisfaction. As can be seen in Figure~\ref{fig:unknown_person}, a rather strict prompt was provided which, at first consideration, might suggest an impossible path forward. Still, the advice given was both prudent and sound, and importantly: achievable.  

Furthermore, while it has been truncated from the figure for reasons of brevity, the following advice was also provided in the same response:

\textbf{``Enhancing Security Measures}\\
\begin{enumerate}
    \item \textbf{Physical Access Controls}: Simple measures like locks on doors and windows can significantly enhance security.
    \item \textbf{Visitor Management}: A basic visitor management policy, including sign-in and possibly accompanied visits for unknown visitors.
    \item \textbf{Community Vigilance}: Engage with neighboring businesses or buildings to report suspicious activity.
    \item \textbf{Security Assessments}: Periodically assess your physical security measures to identify and rectify weaknesses.''
\end{enumerate}

Each of these recommended items can themselves be made a SMART-compliant set of goals to further enhance the physical security posture of an organization. 

\subsection{Relevance}
In building an IRP and associated SOPs, a key factor is ensuring the relevance of the work. Within the information security space (as well as software development in general), a phenomenon often referred to as ``bike-shedding'' arises, originating from the ``law of triviality''~\cite{Parkinson1986-hl}. This notion suggests that individuals may sometimes disproportionately focus on issues of objectively lesser concern than is appropriate for the scope of a given set of tasks or projects. The term is derived from a hypothetical scenario in which an organization responsible for constructing a nuclear power plant spends excessive time deliberating over the color of the employee bike shed rather than focusing on the critical aspects of the plant's safety and functionality.

To mitigate or eliminate bike-shedding in incident response planning and ensure that the work remains of high value and relevance, LLMs can assist by prioritizing through gap analysis. Specifically, an LLM can analyze a list of existing SOPs alongside names of forthcoming candidate SOPs to help identify those of the highest value in the near term. This assessment is not limited to a technical perspective but extends to regulatory and compliance considerations as well. This approach is particularly effective when the context has been set regarding which regulations are applicable to the business, allowing the LLM to tailor its recommendations to ensure compliance and address critical security needs.

\subsubsection{Utilizing LLMs for Strategic Prioritization}
LLMs can provide invaluable assistance in identifying and prioritizing the development of IRPs and SOPs by:
\begin{itemize}
    \item Analyzing current cybersecurity posture of the organization and identifying gaps where additional SOPs could fortify the organization's defenses.
    \item Reviewing the latest cybersecurity threats and trends to recommend updates to existing SOPs or the creation of new ones that address emerging risks.
    \item Considering the regulatory landscape to ensure that all recommended SOPs align with compliance requirements, thereby minimizing legal and financial risks.
\end{itemize}

Furthermore, LLMs can facilitate stakeholder engagement by synthesizing complex security and compliance information into digestible, executive-level summaries. This capability enhances decision-making processes by providing clear, concise data on which aspects of incident response planning should be prioritized to align with business objectives and regulatory mandates.

\subsubsection{Continuous Improvement and Adaptation}
The dynamic nature of cybersecurity threats necessitates a continuous improvement approach to IRP and SOP development. LLMs can support this by:
\begin{itemize}
    \item Regularly reassessing the organization's security posture and SOP relevance in light of new information, ensuring that incident response planning remains agile and effective.
    \item Facilitating a feedback loop from incident response exercises and real-world incidents to refine SOPs and IRP strategies continually.
\end{itemize}

By integrating LLMs into the process of developing and maintaining IRPs and SOPs, organizations can ensure that their incident response planning is not only relevant and compliant but also adaptable to the evolving cybersecurity landscape.

\subsection{Time-bound}
The cybersecurity landscape ever changing, and new threats and vulnerabilities are constant as are techniques and tactics used by threat actors. In this context, the swift identification and development of IRPs and SOPs for new scenarios and new technologies is absolutely critical. Strong SOPs should be broadly applicable and reusable, but as unexpected responses are needed, an SOP supported by deep analysis should be written (e.g., an SOP for disconnecting a campus from the Internet). This section explores the criticality of setting and adhering to time-bound objectives in crafting IRPs and SOPs, with a focus on leveraging LLMs to streamline this process.

\subsubsection{Setting Realistic Timelines}
A thorough assessment of an organization's existing incident response capabilities and resources lays the groundwork for realistic timeline setting. Understanding the starting point is essential for accurate planning and expectation management. LLMs excel in identifying gaps within existing IRPs and SOPs by analyzing vast amounts of data and comparing them with current best practices and compliance standards. This capability allows organizations to identify gaps within the procedures on which new procedures are being built. This helps identify antecedent tasks that may be needed before those under consideration can be built. 

LLMs can dramatically reduce the time required to draft comprehensive IRPs and SOPs. By inputting organizational specifics and compliance requirements, LLMs can generate initial drafts that teams can refine, significantly accelerating the development process.

\subsubsection{Incorporating Time-bound Goals into IRP/SOP Development}
Establishing specific milestones within the IRP and SOP development process is crucial. For instance, completing the draft of an IRP within four weeks or finalizing a set of SOPs by the end of the quarter provides clear targets for the team to aim for. Furthermore, allocating time for feedback and revisions is essential for an iterative approach aiming to optimize the plan. Setting clear deadlines for each review phase ensures that the development process remains on track, allowing for iterative improvements without derailing the overall timeline.

The unexpected can occur in the security space, which may negatively impact the timeline. To reduce the risk of missing a deadline due to unplanned higher-priority tasks, scheduling regular progress reviews helps monitor the advancement of IRP and SOP development against the established timeline. These check-ins facilitate timely adjustments to the plan, ensuring that the development process remains aligned with the set objectives.

\subsection{Post-mortems \& Retrospectives}
As an incident unfolds, maintaining detailed records is paramount not only for evidence preservation but also for a comprehensive post-incident analysis. These records serve multiple purposes, such as elucidating communication dynamics and decision-making processes during the incident.

Following an incident, organizations commonly conduct a ``Post-mortem'' or ``Retrospective'' meeting—terms that, while sometimes used interchangeably, may convey nuanced differences depending on the operational framework. For clarity and simplicity, we will treat them as synonymous in this context. These sessions are crucial for dissecting actions taken, identifying areas for improvement, and enhancing future response efforts. For instance, resolving ambiguities in decision-making authority or updating SOPs based on technical challenges encountered are typical outcomes of these meetings.

Consider the scenario where an organization responds to a stolen device report. The partial incident log (Figure~\ref{fig:stolen_device_partial_log}) captures the sequence of actions taken by the security team, led by an individual named Bob, from the initial report to the execution of various SOPs aimed at mitigating the incident's impact.

\begin{figure}
    \centering
    \small
    \begin{tcolorbox}[colback=gray!10,colframe=black,sharp corners]
        09:00 AM - Incident reported by Alice. Device identified as stolen. Incident Response Plan for ``Stolen Device'' initiated by the security team lead, Bob.
        \\
        \\
        09:05 AM - Bob reviews the ``Remote Wipe'' SOP to ensure proper execution. The SOP aims to remotely erase all sensitive data from the stolen device to prevent unauthorized access.
        \\
        \\
        09:10 AM - Bob attempts to execute the ``Remote Wipe'' procedure. It is discovered that the documentation for the procedure is out of date, and the current remote management platform's interface has changed since the last update.
        \\
        \\
        09:15 AM - After navigating the updated interface with some trial and error, Bob successfully initiates the remote wipe on the stolen device. Confirmation of the wipe initiation is logged.
        \\
        \\
        09:30 AM - Bob proceeds to the ``Reset Password'' SOP to change the credentials associated with the stolen device's user account to prevent unauthorized access from any synchronized services or applications.
    \end{tcolorbox}
    \caption{Stolen Device Partial Incident Log}
    \label{fig:stolen_device_partial_log}
\end{figure}

\begin{figure}
    \centering
    \small
    \begin{tcolorbox}[colback=gray!10,colframe=black,sharp corners]
        \textbf{Prompt:} ``In this incident post-mortem, please add commentary and feedback to the log of events. Provide suggestions for future incidents of the same type. $<$log of incident$>$''\\
        \\
        \textbf{ChatGPT 4 Response:} ``The incident log provides a structured and chronological account of the response to a stolen device, showcasing the execution of the Incident Response Plan (IRP) effectively. Below are some commentary and feedback on the log, along with suggestions for future incidents: 
        \begin{enumerate}
            \item \textbf{Prompt Reporting and Initiation}: The incident was reported and the IRP initiated promptly by the security team lead, demonstrating a quick response to the security incident. This rapid action is crucial in minimizing potential data breaches or unauthorized access associated with the stolen device.
            \item \textbf{Challenges with Outdated SOPs} The incident highlighted a significant issue with the ``Remote Wipe'' Standard Operating Procedure (SOP) being outdated. It's critical for organizations to regularly review and update their SOPs to align with current technologies and platforms. A periodic review schedule should be established, such as bi-annual or annual reviews, to ensure all documentation is up to date.''
        \end{enumerate}
    \end{tcolorbox}
    \caption{Partial Post-mortem analysis}
    \label{fig:postmortem_analysis}
\end{figure}

In Figures~\ref{fig:stolen_device_partial_log} and~\ref{fig:postmortem_analysis} we see a partial record of a stolen device incident. This example log file represents the steps taken by Bob (in the security team) in response to the stolen device reported by Alice.  At the conclusion of the incident, the log file and a brief context was provided to the LLM ChatGPT 4 and asked to provide both a commentary and recommendations for improvement.

We can see that the LLM correctly observed the need to update the SOPs when the procedure changes. Indeed the feedback to the short incident log was extensive and in addition to Figure~\ref{fig:postmortem_analysis}, included many other recommendations including the following:

\begin{itemize}
    \item \textbf{Automated Alerts for SOP Updates}: Implement a system that alerts the responsible team when third-party platforms or internal systems undergo significant updates that could affect SOPs. This could involve monitoring release notes or subscribing to update alerts from service providers.
    \item \textbf{Regular Training and Drills}: Conduct regular training sessions and simulation drills for the incident response team to ensure familiarity with updated SOPs and to practice navigating new interfaces or procedures. This can help reduce the response time in real incidents.
\end{itemize}
 
The integration of LLMs into the post-mortem and retrospective phases of incident response offers a promising avenue for enhancing organizational learning and response effectiveness. By providing detailed analysis, identifying procedural gaps, and suggesting actionable improvements, LLMs can significantly contribute to the development of robust incident response strategies. As this technology continues to evolve, its potential to transform traditional post-incident analysis practices grows, promising a future where organizations are better equipped to learn from and mitigate the impact of security incidents.
\section{Conclusion}
\label{conclusion}

The integration of Large Language Models into cybersecurity teams marks an exciting and transformative phase in incident response planning and execution. By leveraging the vast knowledge and analytical capabilities of LLMs, organizations can rapidly assess threats, devise response strategies, and generate actionable insights, all within a fraction of the time traditionally required. While the creation of dynamic, version-controlled response plans remains an essential task, LLMs significantly alleviate the associated workload, making the process more efficient and effective. Despite the promise of LLMs, it's important to navigate potential challenges, including data privacy and the risk of dependency on automated systems. As LLMs continue to evolve, their potential to revolutionize incident response planning is undeniable, promising not only to enhance operational efficiency but also to foster a more proactive and resilient cybersecurity posture.

\bibliographystyle{acm}
\bibliography{bibliography}

\begin{thebibliography}{1}

\bibitem{nist800_61r2}
{\sc Cichonski, P., Millar, T., Grance, T., and Scarfone, K.}
\newblock Nist special publication 800-61 rev. 2, computer security incident
  handling guide, Aug 2012.

\bibitem{Dolan_1981}
{\sc Dolan, G.~T.}
\newblock There’s a s.m.a.r.t. way to write management’s goals and
  objectives.
\newblock {\em Management Review 70}, 11 (1981), 35–36.

\bibitem{Parkinson1986-hl}
{\sc Parkinson, C. N. c.~N.}
\newblock {\em Parkinson's law}.
\newblock Penguin business library. Penguin Books, Harlow, England, Mar. 1986.

\bibitem{semver}
{\sc Preston-Werner, T.}
\newblock Semantic versioning 2.0.0.

\bibitem{Swanson_Bowen_Phillips_Gallup_Lynes_2010}
{\sc Swanson, M., Bowen, P., Phillips, A.~W., Gallup, D., and Lynes, D.}
\newblock Nist special publication 800-34 rev.1, contingency planning guide for
  federation information systems, 2010.

\end{thebibliography}

\end{document}